\documentclass[aps,prl,floatfix,showpacs,twocolumn]{revtex4-1}
\usepackage{ulem}
\usepackage{dcolumn}
\usepackage{bm}
\usepackage{color}
\usepackage{amssymb}
\usepackage{amsmath}
\usepackage{graphicx}
\usepackage{amsfonts}
\usepackage{slashed}
\usepackage{pstricks}
\usepackage{float}
\usepackage{hyperref}
\usepackage{array}
\allowdisplaybreaks
\usepackage{dcolumn}
\usepackage{epsf}
\usepackage{float}

\begin{document}
\title{Ab initio calculation of neutral-current $\nu$-$^{12}$C inclusive quasielastic
scattering}
\author{
A.\ Lovato$^{\, {\rm a,b} }$,
S.\ Gandolfi$^{\, {\rm c} }$,
J.\ Carlson$^{\, {\rm c} }$,
Ewing\ Lusk$^{\, {\rm d} }$,
Steven\ C.\ Pieper$^{\, {\rm b} }$,
and R.\ Schiavilla$^{\, {\rm e,f} }$
}
\affiliation{
$^{\,{\rm a}}$\mbox{INFN-TIFPA Trento Institute of Fundamental Physics and Applications, 38123 Trento, Italy}\\
$^{\,{\rm b}}$\mbox{Physics Division, Argonne National Laboratory, Argonne, IL 60439}\\
$^{\,{\rm c}}$\mbox{Theoretical Division, Los Alamos National Laboratory, Los Alamos, NM 87545}\\
$^{\,{\rm d}}$\mbox{Mathematics and Computer Science Division, Argonne National Laboratory, Argonne, IL 60439}\\
$^{\,{\rm e}}$\mbox{Theory Center, Jefferson Lab, Newport News, VA 23606}\\
$^{\,{\rm f}}$\mbox{Department of Physics, Old Dominion University, Norfolk, VA 23529}
}
\date{\today}
\begin{abstract}
Quasielastic neutrino scattering is an important aspect of the 
experimental program to study fundamental neutrino properties 
including neutrino masses, mixing angles, the mass hierarchy and
CP-violating phase.  
Proper interpretation of the experiments requires reliable theoretical calculations
of neutrino-nucleus scattering.
In this paper we present calculations of response
functions and cross sections by neutral-current scattering of neutrinos off $^{12}$C.
These calculations are based on realistic treatments of nuclear interactions
and currents, the latter including the axial-, vector-, and vector-axial 
interference terms crucial for determining the difference between
neutrino and anti-neutrino scattering and the CP-violating phase.
We find that the strength and energy-dependence of two-nucleon
processes induced by correlation effects and interaction currents
are crucial in providing the most accurate description of
neutrino-nucleus scattering in the quasielastic regime.
\end{abstract} 
\index{}\maketitle

Many accelerator experiments
are running~\cite{mb_web,nova_web,t2k_web,minerva_web},
or are being planned~\cite{dune_web,hk_web}, 
to measure neutrino oscillation parameters including 
the neutrino mass hierarchy and 
the charge-conjugation parity (CP) violating phase
that differentiates the oscillation probabilities of 
neutrinos ($\nu$) and anti-neutrinos ($\overline\nu$).
These experiments employ nuclear targets like
$^{16}$O (T2K) or $^{56}$Fe and $^{208}$Pb (MINER$\nu$A)
or $^{40}$Ar (DUNE), and use event generators (EGs)~\cite{Alvarez-Ruso:2017oui}
to analyze the scattering data by
modeling the nucleus and reaction mechanisms.  The EGs also
provide information on key features of the experiment (signal and
background event-rate distributions, systematic errors, {\it etc.})
crucial for the interpretation of the data in terms of oscillation parameters.
Even in an experiment involving both near and far detectors, 
the EGs and associated nuclear physics models
are required to determine the neutrino energy in order
to perform analyses where the ratio of length to energy $(L/E)$ is a 
critical input.

A large and growing body of work highlights the sensitivities of experimental analyses to
systematic uncertainties in the nuclear physics~\cite{Ankowski:2016bji,Mosel:2016cwa,Coloma:2013tba}.
Phenomenological approaches exist (see Refs.~\cite{Benhar:2015wva,Katori:2016yel}) 
using electron scattering data as constraints, but they often incorporate significant approximations and 
are more directly tested in vector-current processes.  For example, it is known that
two-nucleon currents and correlations also play an important role
in axial-current matrix elements~\cite{Lovato:2014},
particularly in the axial-vector interference ones that
determine the differences between $\nu$ and $\overline\nu$ cross sections.

In the present paper, we report on an {\it ab initio} quantum Monte Carlo (QMC) calculation,
based on Green's function Monte (GFMC) methods~\cite{Carlson:2015}, of $\nu$-$^{12}$C inclusive scattering
induced by neutral-current interactions.  While limited in kinematical scope to the quasielastic
region, it has nevertheless the advantage of relying on a first-principles
description of nuclear dynamics.  In such a description, the nucleons interact with each other via
effective two- and three-body potentials---respectively, the Argonne $v_{18}$
(AV18)~\cite{Wiringa:1995} and Illinois-7 (IL7)~\cite{Pieper:2001} models---and with
electroweak fields via effective currents, including one- and two-body terms~\cite{Shen:2012}.
The GFMC methods then allow us to fully account, without approximations, for
the complex many-body, spin- and isospin-dependent correlations induced by these
nuclear potentials and currents, and for interaction effects in the final nuclear states~\cite{Carlson:1992}.
For moderate momentum transfers and quasielastic energy transfers, the results of these
calculations should provide a useful benchmark for testing predictions from EGs and/or
approaches based on approximate schemes of nuclear dynamics.  
 
A recent GFMC calculation of the $^{12}$C longitudinal and transverse
electromagnetic response functions~\cite{Lovato:2016gkq} is in very satisfactory agreement
with experimental data, obtained from Rosenbluth separation of inclusive
$(e,e^\prime)$ cross sections. This agreement validates the dynamical framework
and, in particular, the model for the vector currents adopted here.  An interesting outcome
of this study is the realization that interaction effects in the final nuclear states
and two-body terms in these currents substantially affect the distribution
of strength in the response as a function of the energy transfer $\omega$.  
Final-state
interactions (and initial- and final-state correlations) shift strength away from the quasielastic
region into the threshold and high $\omega$ regions in both the longitudinal and transverse
response functions.  Two-body current contributions, though, while
negligible in the longitudinal response, significantly 
increase the transverse one
in the quasielastic peak, thus off-setting the quenching.

The energy dependence of the cross section is especially relevant for neutrino experiments,
since it directly impacts the analysis of these experiments in terms of oscillation parameters and CP-violating phase.  
Earlier studies of integral properties of the response, either
sum rules~\cite{Lovato:2014} or Laplace transforms of the response itself, so called Euclidean
response functions~\cite{Carlson:1992,Lovato:2015}, have indicated that two-nucleon
currents are important.  However,
these properties only provide indirect information on the strength distribution as a function of $\omega$.

\begin{center}
\begin{figure}[h!]
\includegraphics[width=8.36cm]{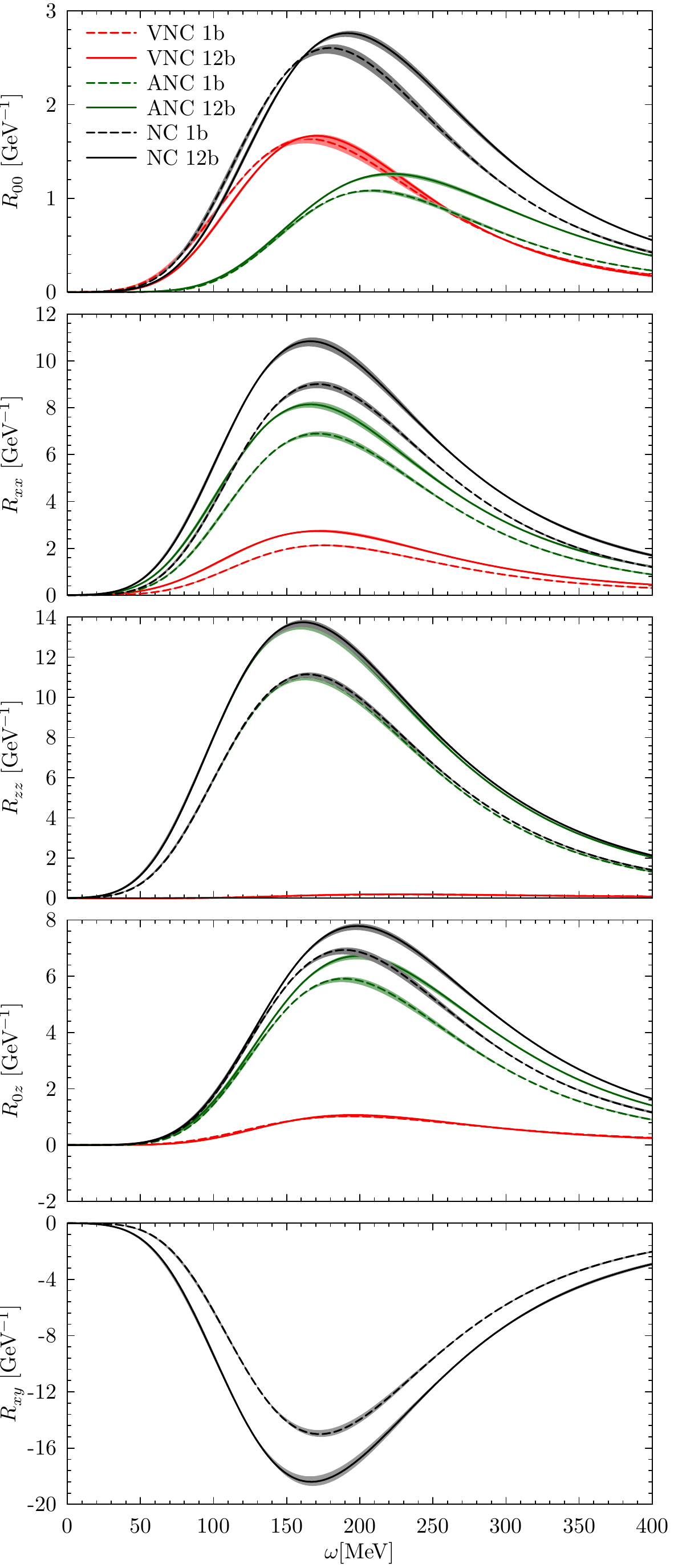}
\caption{(Color online) Neutral-current response functions
in $^{12}$C at momentum transfer $q\,$=$\,$570 MeV/c, corresponding to the AV18/IL7 Hamiltonian and
obtained with one-body only (dashed lines) and one- and
two-body (solid lines) currents. The narrow bands indicate the uncertainty in the maximum-entropy inversion.
The vector and axial contributions are shown separately in all cases but for $R_{xy}$.  See text for further
explanations.}
\label{fig:f1}
\end{figure}
\end{center}

The differential cross section for $\nu$ and $\overline\nu$  inclusive
scattering off a nucleus induced by neutral-weak currents can be expressed as~\cite{Shen:2012}
\begin{align}
\frac{d\sigma}{d\omega \, d\Omega} =& \frac{G_F^2}{2\pi^2} k^\prime E^\prime \cos^2\frac{\theta}{2}
 \Bigg[ R_{00}(q,\omega)+
\frac{\omega^2}{q^2}R_{zz}(q,\omega)\nonumber\\
&-\frac{\omega}{q}R_{0z}(q,\omega)
+ \Big(\tan^2\frac{\theta}{2} + \frac{Q^2}{2\,q^2}\Big)R_{xx}(q,\omega)\nonumber \\
&\mp \tan\frac{\theta}{2} \sqrt{\tan^2\frac{\theta}{2}+
\frac{Q^2}{q^2}}R_{xy}(q,\omega)\Bigg]\ ,
\label{eq:xs}
\end{align}
where $-$ ($+$) refers to $\nu$ ($\overline\nu$),
$k^\prime$ and $E^\prime$ are the momentum and energy of the outgoing neutrino, $q$ and $\omega$ 
are the momentum and energy transfers with $Q^2=q^2 -\omega^2$ being
the four-momentum transfer, $\theta$ is the outgoing neutrino scattering angle relative
to the incident neutrino beam direction, and $G_F=1.1803\times 10^{-5}$ GeV$^{-2}$
as obtained from an analysis of super-allowed $0^+\to 0^+$ $\beta$-decays~\cite{Towner:1999}.

The nuclear response functions are schematically given by
\begin{align}
R_{\alpha\beta}(q,\omega) &=\sum_f \langle f | j^{NC}_\alpha({\bf q},\omega) |0\rangle 
\langle f | j^{NC}_\beta({\bf q},\omega) |0\rangle^* \nonumber\\
& \times \delta(E_f-\omega-E_0)\ ,
\label{eq:res_def}
\end{align}
where $|0\rangle$ and $|f\rangle$ represent the nuclear initial ground-state and final bound-
or scattering-state of energies $E_0$ and $E_f$, and $j^{NC}_\alpha({\bf q},\omega)$ 
denotes the appropriate components of the weak neutral current ($NC$).
Explicit expressions for these currents and response functions are listed
in Ref.~\cite{Shen:2012}; here, it suffices to note that
the subscripts $0$ and $z$ refer to, respectively, the charge $\rho^{NC}$
and longitudinal component of the current ${\bf j}^{NC}$, and $x$ and $y$ to
the transverse components of ${\bf j}^{NC}$.  The momentum transfer ${\bf q}$
is taken along the spin quantization axis---the $z$ axis.

The calculation of the response functions proceeds along similar lines to 
that of Ref.~\cite{Lovato:2016gkq}.
We compute the Laplace transforms of $R_{\alpha\beta}(q,\omega)$ with respect to
$\omega$ which reduce to
the following current-current correlators
\begin{align}
E_{\alpha\beta}(q,\tau)\!&=\!\langle0| j^{NC\,\dagger}_\alpha({\bf q},\omega_{\rm qe}) e^{-(H-E_0)\tau} 
j^{NC}_\beta({\bf q},\omega_{\rm qe}) |0\rangle\nonumber\\
&-|F_{\alpha\beta}(q)|^2 {\rm e}^{-\tau \omega_{\rm el}} \ ,
\label{eq:euc_me}
\end{align}
where $H$ is the nuclear Hamiltonian and the elastic contributions proportional to
the (elastic) form factors $F_{\alpha\beta}(q)$ have been removed ($\omega_{\rm el}$
is the energy of the recoiling ground state).  The energy dependence of ${\bf j}^{NC}(q,\omega)$
comes in via the nucleon and nucleon-to-$\Delta$ transition weak neutral form factors, which are
functions of $Q^2$.  We freeze the $\omega$ dependence by fixing $Q^2$ at the value
$Q^2_{\rm qe}=q^2-\omega_{\rm qe}^2$ with the quasielastic energy transfer $\omega_{\rm qe}$
given by $\omega_{\rm qe}=\sqrt{q^2+m^2}-m$ ($m$ is the nucleon mass).  This is needed in
order to exploit the completeness over the nuclear final states in evaluating the Laplace transforms
of $R_{\alpha\beta}(q,\omega)$~\cite{Lovato:2016gkq}.   Lastly, since terms in the states $j_\alpha^{NC}|0\rangle$
involve gradients of the ground-state wave function~\cite{Shen:2012}, we evolve, rather
than the exact ground state $|0\rangle$, our best variational state $|0_{\,T}\rangle$
in order to reduce the computational cost.  Comparison between sum-rule results obtained
with either $|0\rangle$ or $|0_{\,T}\rangle$ indicates that this is an excellent approximation~\cite{Lovato:2013}.

Evaluations of the various correlators are carried out in two steps. First, an 
unconstrained imaginary-time propagation of $|0_{\,T}\rangle$ is performed 
and stored. Next, the states $j^{NC}_\beta({\bf q},\omega_{\rm qe}) |0_{\,T}\rangle$ are
evolved in imaginary-time following the path previously saved.  During this
evolution scalar products of $e^{-(H-E_0)\tau_i}  j^{NC}_\beta({\bf q},\omega_{\rm qe}) 
|0_{\,T}\rangle$ with $j^{NC}_\alpha({\bf q},\omega_{\rm qe})|0_{\,T}\rangle$
are computed on a grid of $\tau_i$ values, and from these scalar products estimates
of $E_{\alpha\beta}(q,\tau_i)$ are obtained (a more extended discussion of the methods
is in Refs.~\cite{Carlson:1992,Carlson:2002}).  The computer programs are written
in FORTRAN and use MPI and OPENMP for parallelization.  While Monte Carlo
calculations are thought of as ``embarrassingly parallel'', the GFMC propagation involves
killing and replication of configurations which in fact could lead to significant inefficiencies in a parallel
environment---in the present case, the Mira supercomputer of the Argonne Leadership
Computing Facility.  Moreover, for a nucleus such as $^{12}$C with
its large number (about $4 \times 10^6$) of spin and isospin states, the calculation
of a single Monte Carlo sample
must be spread  over many nodes.  For these reasons, the Asynchronous Dynamic Load Balancing
(ADLB) and Distributed MEMory (DMEM) libraries~\cite{Lusk:2010}, which operate under MPI, were
developed.  As a consequence, parallelization efficiency close
to $95\%$ is achieved using 8192 nodes of Mira.  Approximately one million node hours
were used for the calculation reported here.

Maximum entropy techniques, developed specifically for the present problem
in Ref.~\cite{Lovato:2016gkq}, are then utilized to perform the analytic continuation
of the Euclidean response functions, corresponding to the ``inversion'' of the Laplace
transforms.  The resulting $^{12}$C neutral
weak response functions $R_{\alpha\beta}(q,\omega)$ are displayed in Fig.~\ref{fig:f1}
for $q\,$=$\,$570 MeV/c.  In this connection, we note that the presence
of low-lying excitations of the $^{12}$C nucleus complicates the determination of
$R_{\alpha\beta}(q,\omega)$ for $\omega$ values near threshold~\cite{Lovato:2016gkq}.
Resolving the corresponding peaks would require imaginary-time evolution
to $\tau$ values of the order of $1/\Delta E$,
where $\Delta E$ are the excitation energies of these states ($\Delta E\,$=$\,$ 4.44 MeV
for the lowest $2^+$ excited state in $^{12}$C); due to the Fermion sign problem, this
is not possible. 
Each of these peaks, however,
is proportional to the square of weak neutral transition form factors
between the ground and relevant excited state.  Because of the rapid
fall-off of these form factors with increasing momentum transfer
this issue is not expected to be of any concern at the relatively high
momentum transfer of interest here.  Of course, these
considerations remain valid for the elastic contributions
alluded to earlier in Eq.~(\ref{eq:euc_me}).

Figure~\ref{fig:f1} shows that contributions from two-body terms in the $NC$ significantly
increase (in magnitude) the response functions obtained in impulse approximation ({\it i.e.}, with
one-body currents) over the whole quasielastic region, except for $R_{00}$ at low $\omega$.
This enhancement is mostly due to constructive interference
between the one- and two-body current matrix elements, and is consistent with that
expected on the basis of sum-rule analyses~\cite{Lovato:2014}. Counter to the electromagnetic
case~\cite{Lovato:2016gkq}, we find that two-body terms in the weak neutral charge produce
substantial excess strength in $R_{00}$ and $R_{0z}$ beyond the quasielastic peak.
In the $00$, $0z$, $zz$, and $xx$ response functions the vector ($VNC$) and axial ($ANC$)
components of the weak neutral current, $j^{NC}_\alpha\,$=$\,j^{VNC}_\alpha
+j^{ANC}_\alpha$, do not interfere; in these cases, $R_{\alpha\beta}\,$=$\,R^{VNC}_{\alpha\beta}+
R^{ANC}_{\alpha\beta}$ and the $R_{\alpha\beta}^{VNC}$ and $R_{\alpha\beta}^{ANC}$ are
illustrated separately in Fig.~\ref{fig:f1}.  By contrast, the $xy$ response function
arises solely on account of this interference.  The $ANC$ contribution to
$R_{\alpha\beta}$ is typically much larger than the $VNC$ one (for example,
$R^{ANC}_{xx} \,$$\simeq$$\, 3\times R^{VNC}_{xx}$), except for the charge response
$R_{00}$.  Furthermore, in $^{12}$C the $00$ and $xx$ $VNC$ response functions are
roughly proportional to the longitudinal and transverse electromagnetic response functions
$R_L$ and $R_T$, namely $R_{00/xx}^{VNC} \simeq R_{L/T}/4$.  This is because the isoscalar and
isovector pieces in $j^{VNC}$ are related to the corresponding ones in the electromagnetic
current $j^{EM}$ by the factors, respectively, $-2\,{\rm sin}^2\theta_W$ and
$(1-2\,{\rm sin}^2\theta_W$)  (${\rm sin}^2\theta_W \simeq 0.23$), and the matrix
elements of these pieces add up incoherently in the response of an isoscalar target
such as $^{12}$C.
\begin{center}
\begin{figure}[!h]
\includegraphics[width=8.35cm]{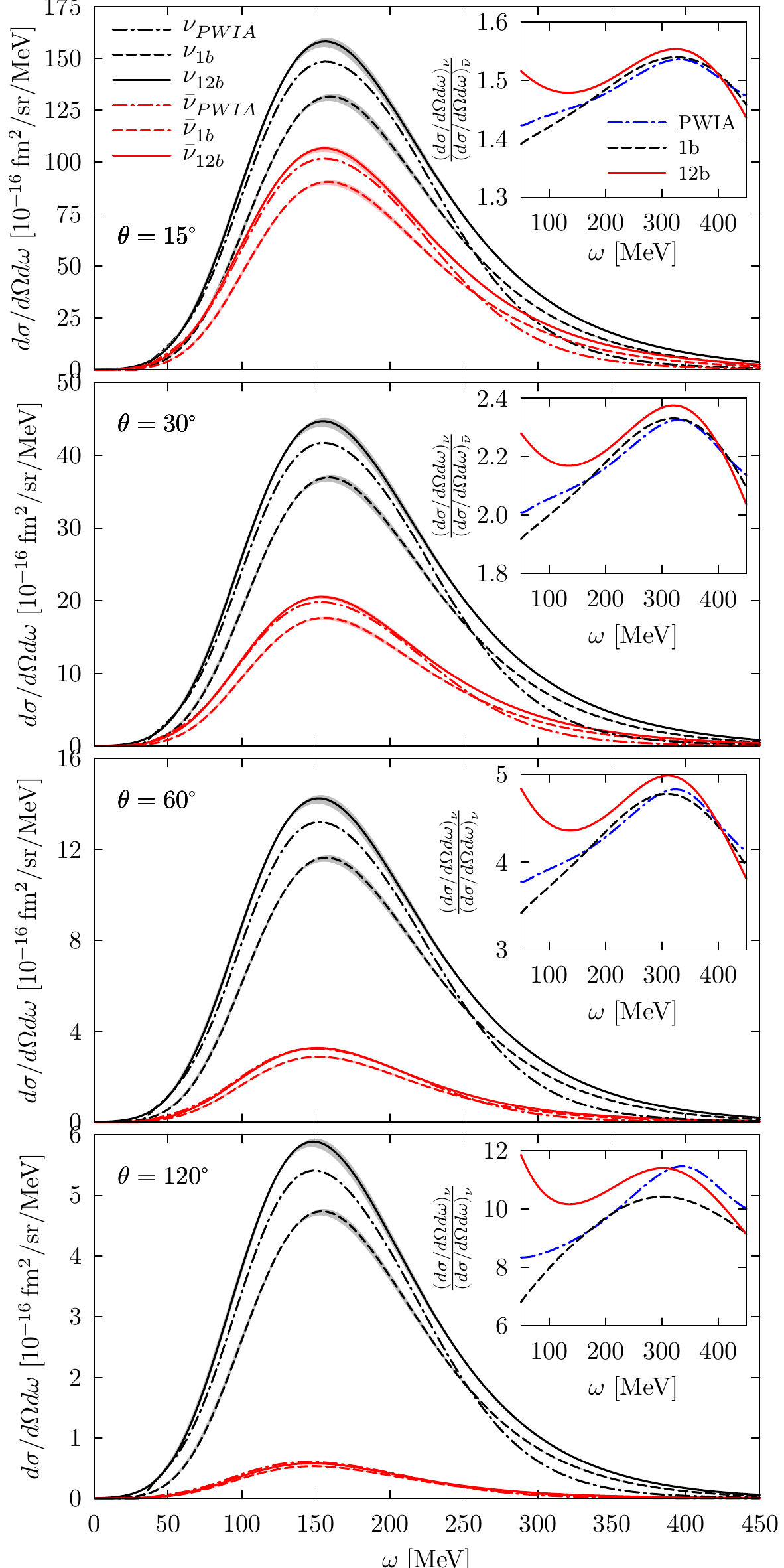}
\caption{(Color online) Weak neutral $\nu$ (black curves) and $\overline{\nu}$ (red curves)
differential cross sections in $^{12}$C at $q\,$=$\,$570 MeV/c, obtained with one-body only and
one- and two-body terms in the $NC$.  The final neutrino angle is indicated in each panel.
The insets show ratios of the $\nu$ to $\overline{\nu}$ (central-value) cross sections.  Also shown are the
PWIA results. }
\label{fig:f2}
\end{figure}
\end{center}

The two-body terms in the $ANC$ increase the
one-body $R^{ANC}_{xx}$ response by about 20\% in the quasielastic region.  
This increase is much larger than the $\simeq\,$2--4\% obtained 
in the case of Gamow-Teller rates between low-lying states near 
threshold, induced by the axial component of the weak 
charged current~\cite{Pastore:2017uwc}. In those calculations a significant
reduction of the relevant matrix elements arose from
nuclear correlations, which are also included here.  It would be very intriguing to
study the interplay between, and evolution of, correlation effects and two-body current contributions
as the momentum and energy transfers increase from the threshold
regime of relevance in discrete transitions between low-lying states,
to the intermediate regime ($\sim\,50$--100 MeV) of interest to neutrino
scattering in astrophysical environments or 
neutrinoless double beta decay~\cite{Engel:2016,Pastore:2017bb},
to the quasielastic regime being studied here.

In Fig.~\ref{fig:f2} we show the $\nu$ and $\overline\nu$ differential cross sections
and the $\nu / \overline\nu$ ratios for a fixed value of the three-momentum transfer as
function of the energy transfer for a number of scattering angles.  In terms of these
variables, the initial energy $E$ of the neutrino is given by
\begin{equation}
E =\frac{\omega}{2} \left[ 1+\sqrt{1+\frac{Q^2}{\omega^2\,{\rm sin}^2(\theta/2)} }\,\right]\ ,
\end{equation}
and its final energy $E^\prime\,$=$\, E-\omega$: for example, at $\theta\,$=$\,$15$^\circ$ the initial
energy decreases from 2.2 GeV to 1.6 GeV as $\omega$ increases from threshold to 450 MeV;
at $\theta\,$=$\,$120$^\circ$ the initial energy increases from roughly 0.3 GeV to slightly over 0.5 GeV as $\omega$
varies over the same range.  Thus the present results computed at fixed $q\,$=$\,$570 MeV/c as a function of $\omega$
span a broad kinematical range in terms of $E$ and $E^\prime$---the kinematical variables
most relevant for the analysis of accelerator neutrino experiments.   

Because of the cancellation in Eq.~(\ref{eq:xs})
between the dominant contributions proportional to the $R_{xx}$ and $R_{xy}$ response functions,
the $\overline{\nu}$ cross section decreases rapidly relative to the $\nu$ cross section as the scattering angle changes
from the forward to the backward hemisphere.  For this same reason, two-body current contributions
are smaller for the $\overline{\nu}$ than for the $\nu$ cross section, in fact
becoming negligible for the $\overline{\nu}$ backward-angle cross section.
As the angle changes from the forward to the backward hemisphere, the
$\nu$ cross section drops by almost an order of magnitude, and in the limit
$\theta\,$=$\,$ 180$^\circ$ is just proportional to $R_{xx}(q,\omega)-R_{xy}(q,\omega)$.

 For comparison, we also show results obtained in the plane-wave impulse 
approximation (PWIA), in which only one-body currents are retained 
$j^{NC}_\alpha = \sum_i j_{\alpha}^{NC}(i)$. In PWIA the struck nucleon with initial momentum 
$\mathbf{p}$ absorbs the external field momentum $\mathbf{q}$ and transitions to a particle
state of momentum $\mathbf{p+q}$ without further interactions with the spectator nucleons.  In its simplest formulation, the PWIA response functions are
\begin{align}
R_{\alpha\beta}^{PWIA}&({\bf q},\omega)=\int \frac{d^3p}{(2\pi)^3} n(\mathbf{p}) r_{\alpha\beta}(\mathbf{p},\mathbf{q},\omega)\nonumber\\
&\times \delta\Big[\omega-\overline{E}-\frac{(\mathbf{p+q})^2}{2m}-\frac{p^2}{2(A-1)m}\Big]\, .
\end{align}
In the above equation $\overline{E}$ is the average removal energy, $n(\mathbf{p})$ is the momentum distribution
of the struck nucleon, which we take from~\cite{Wiringa:2013ala}, and the single-nucleon 
coupling to the external neutral-current field is given by
 \begin{align}
 r_{\alpha\beta}(\mathbf{p},\mathbf{q},\omega)&=\frac{1}{4}\sum_{\eta,\eta^\prime}\langle\mathbf{p}+\mathbf{q},\eta^\prime|  j_{\alpha}^{NC}(1) | \mathbf{p},\eta\rangle \nonumber\\
&\times \langle\mathbf{p},\eta|  j_{\beta}^{NC}(1)| \mathbf{p+q},\eta^\prime\rangle\, ,
 \end{align}
where $\eta$ ($\eta^\prime$) indicates the spin-isospin state of the initial (final) nucleon. 
The comparison between the PWIA and the one-body GFMC curves
highlights how correlations and final state interactions quench the quasielastic peak, redistributing 
strength to the threshold and high-energy transfer regions.

In summary, we find substantial two-nucleon contributions to the neutral-current scattering
of neutrinos and anti-neutrinos from $^{12}$C. These contributions are significant over the entire
quasielastic region, and are very important in each of the vector, axial, and
axial-vector interference response functions.  They significantly impact the magnitude of
the cross sections, their energy dependence, and particularly the ratio of neutrino to anti-neutrino
cross sections.  It will be important to compare different, more approximate
treatments of $\nu$-$A$ scattering to these calculations, and also to extend the present
calculations over a wider range of energy and momentum transfers.

\acknowledgments
Conversations and e-mail exchanges with M. Martini, J. Nieves, and
N. Rocco are gratefully acknowledged.
This research is supported by the U.S.~Department of Energy, Office
of Science, Office of Nuclear Physics, under contracts DE-AC02-06CH11357
(A.L., E.L.~and S.C.P.), DE-AC52-06NA25396 (S.G.~and J.C.), DE-AC05-06OR23177
(R.S.), and by the NUCLEI SciDAC and LANL LDRD programs.
Under an award of computer time provided by the INCITE program,
this research used resources of the Argonne Leadership Computing
Facility at Argonne National Laboratory, which is supported by the
Office of Science of the U.S.~Department of Energy under contract
DE-AC02-06CH11357.  It also used resources 
provided by Los Alamos Open Supercomputing, by the Argonne LCRC,
and by the National Energy Research Scientific Computing Center,
which is supported by the Office of Science of the U.S.~Department
of Energy under contract DE-AC02-05CH11231.

\bibliography{biblio}

\end{document}